\documentclass[12pt]{article}

\usepackage[T1]{fontenc}
\usepackage[utf8]{inputenc}
\usepackage{authblk}
\usepackage{graphicx}
\usepackage{mathtools}
\usepackage{url}

\let\origthanks\thanks
\renewcommand\thanks[1]{\begingroup\let\rlap\relax\origthanks{#1}\endgroup}

\title{Efficient Bayesian estimation of the generalized Langevin equation from data}

\author{Clemens Willers\thanks{clemens.willers@uni-muenster.de, ORCID ID: 0000-0001-5777-0514}}
\author{Oliver Kamps\thanks{okamp@uni-muenster.de, ORCID ID: 0000-0003-0986-0878}}
\affil{Center for Nonlinear Science (CeNoS), Westf{\"a}lische Wilhelms-Universit\"at M\"unster, Corrensstr.\ 2, 48149 M\"unster, Germany}
\date{\today}

\begin{document}

\maketitle

\begin{abstract}
Modeling non-Markovian time series is a recent topic of research in many fields such as climate modeling, biophysics, molecular dynamics, or finance. The generalized Langevin equation (GLE), given naturally by the Mori-Zwanzig projection formalism, is a frequently used model including memory effects. In applications, a specific form of the GLE is most often obtained on a data-driven basis. Here, Bayesian estimation has the advantage of providing both suitable model parameters and their credibility in a straightforward way. It can be implemented in the approximating case of white noise, which, far from thermodynamic equilibrium, is consistent with the fluctuation-dissipation theorem. However, the exploration of the posterior, which is done via Markov chain Monte Carlo sampling, is numerically expensive, which makes the analysis of large data sets unfeasible. In this work, we discuss an efficient implementation of Bayesian estimation of the GLE based on a piecewise constant approximation of the drift and diffusion functions of the model. In this case, the characteristics of the data are represented by only a few coefficients, so that the numerical cost of the procedure is significantly reduced and independent of the length of the data set. Further, we propose a modification of the memory term of the GLE, leading to an equivalent model with an emphasis on the impact of trends, which ensures that an estimate of the standard Langevin equation provides an effective initial guess for the GLE. We illustrate the capabilities of both the method and the model by an example from turbulence.
\end{abstract}

\vspace{2pc}

\section{Introduction\label{intro}}

Time series of complex dynamical systems often show stochastic behavior. In many cases these can be modeled in terms of the standard Langevin equation (SLE). This model includes the assumption that the process fulfills the Markov property, i.e., that it has no memory. If it is not possible to derive an SLE directly from the physical mechanisms that define the dynamics of the system, it can be estimated from data \cite{directEstFriedrich,directEstSiegert,DEcorrection0,DEcorrection1,DEcorrection2}, which has been done for various systems from different fields of science \cite{ReviewFriedrichPeinke}. 

However, if an observed process includes memory effects, a standard Langevin description is not sufficient. This issue is a recent topic of research and highly relevant, especially in the context of coarse-grained variables (macroscopic variables arising from a so-called coarse-graining procedure which represent effective dynamics in reduced dimensions), if essential and non-essential degrees of freedom cannot be separated in a strict manner. Examples can be found in molecular dynamics (MD) \cite{GLE_PRE,GLE_MD_1,GLE_Jung2,GLE_Lei2,Review_GLE_MD}, climate modeling \cite{GLE_climate,GreenlandIce,closureModels,GLE_ML,DiffModels_withMemory_2020,confpaper_GLE_climate}, biophysics \cite{GLE_PRX}, or finance \cite{GLE_ML,DiffModels_withMemory_2020,GLE_financial,DetermineMZcoeff}, to name a few. In these cases one has to turn to the generalized Langevin equation (GLE) that includes a memory kernel $\mathcal{K}$:

\begin{align}\label{modelGLEoriginal}
\dot{X}_t = D^{(1)}(X_t) + \int_{s=0}^t \mathcal{K}_s X_{t-s}\, ds + \sqrt{D^{(2)}(X_t)}\,\zeta_t.
\end{align}

\noindent Hereby $D^{(1)}$ and $D^{(2)}$ are called drift and diffusion function, respectively, and have arbitrary shape. In thermodynamic equilibrium, the fluctuation-dissipation theorem (FDT) implies that the model contains colored noise $\zeta_t$ which is linked to the memory kernel through $\langle \zeta_t\zeta_{t'} \rangle \propto \mathcal{K}_{t-t'}$. Yet, within the scope of this work, we discuss a GLE that incorporates white noise which is independent of the memory term, i.e., we replace $\zeta_t$ by $\eta_t$ with $\langle \eta_t\eta_{t'} \rangle \propto  \delta(t-t')$. Far from equilibrium, this modification causes no contradiction to the FDT (as pointed out in, e.g., reference \cite{GLE_nonEquilibrium}). An example of this concerning turbulence is presented in section \ref{example}. Interestingly, there is even an example from the field of molecular dynamics in which a GLE with white noise is employed \cite{GLE_PRE}. Also in the case that the Liouville operator of the system is non-Hermitian, there is generally no contradiction in using white noise \cite{nonHermitianLiouvillesOperator}. In other cases, it might still be a reasonable approximation (as assumed in reference \cite{GreenlandIce}), or the model could be considered in a phenomenological manner. Despite the restriction to white noise, we still deal with a very general model class representing systems with memory.

In principle the GLE can be derived systematically through the Mori-Zwanzig formalism by coarse-graining microscopic equations \cite{GrabertsBuch}. However, as for the SLE, a direct derivation is often not accessible. In this case, one aims to estimate the model solely based on measured or simulated data. Here, different approaches can be found in the literature. To begin with, in the case of a linear drift and a constant diffusion function, the GLE can be derived from the autocorrelation function (ACF) of the measured process through solving a Volterra integro-diffential equation \cite{DiffModels_withMemory_2020,DetermineMZcoeff}. Further, various techniques have been developed in the field of MD (for an overview cf. reference \cite{Review_GLE_MD}), where the drift term (which is a conservative force in this context) and the memory kernel are usually estimated in separate steps. The most challenging part is the reconstruction of the memory kernel (which, together with temperature, also determines the noise term in the case of MD in thermodynamical equilibrium). Two important methods were proposed by Jung et. al. \cite{GLE_Jung2,GLE_Jung1} (an iterative procedure) and Lei et. al. \cite{GLE_Lei2,GLE_Lei1} (an approximate description in the Laplace domain). We also mention reference \cite{GLE_BayesianOPTIMIZATION} in which Bayesian optimization is employed to obtain a memory kernel in such a way that the ACF of the model fits the measured one. Examples for the estimation of the force term are listed in \cite{GLE_Lei1}. Moreover, the GLE is approximated by various models whose estimation is simpler: autoregressive (AR) or autoregressive moving-avarage (ARMA) models \cite{GLE_PRE,linearGLE_ARMA}, neural networks \cite{GLE_ML}, an extended SLE including hidden variables \cite{MLE_GLE_MD} (here, the estimation is performed through the expectation-maximization algorithm), or EMR models (empirical model reduction) \cite{closureModels} which are obtained via regression. The latter technique is also applied in reference \cite{GreenlandIce} to estimate a GLE with white noise. There, it is also shown how to perform maximum likelihood estimation (MLE) of the same model. MLE is further used in reference \cite{GLE_PRX} in the context of a specific GLE model belonging to a particular second-order SLE.

Beyond that, the method of Bayesian estimation provides a simple and flexible framework with the advantage that both estimates and their credibility can be determined from a single training data set (cf. section \ref{sec:Bayes}). For the GLE, its implementation is feasible in the case of white noise. It is a straightforward extension of MLE (see above). Further, it is discussed in detail in reference \cite{POD_nonMarkovGLEtype}. However, the exploration of the posterior distribution via Markov chain Monte Carlo (MCMC) sampling, which is necessary for the calculation of credible intervals for estimated parameters, is numerically expensive. Thus, the analysis of large data sets with a length of, e.g., $10^7$ data points is impracticable. In this work, we discuss an efficient implementation of Bayesian estimation of a GLE model incorporating white noise as introduced above. It is facilitated by a piecewise constant approximation of the drift and diffusion functions which allows to represent the characteristics of the data by only a few coefficients. This approach (which has previously been described by Kleinhans for the SLE \cite{KLEINHANS_MLE}) significantly reduces the number of floating point operations (FLOPS) performed during estimation.

In detail, we proceed as follows. First, we propose a reformulation of the memory kernel (leading to an equivalent model, cf. section \ref{sec:reformulation}) such that an estimate of an SLE model can be used as an effective initial guess for the estimation of the GLE. Next, we introduce the concept of Bayesian estimation in more detail (cf. section \ref{sec:Bayes}), derive the posterior of the considered model via Euler-Maruyama approximation (cf. section \ref{sec:likelihood}), and formulate its efficient implementation (cf. section \ref{sec:efficient}). Afterwards, we discuss the generation of an initial guess (cf. section \ref{sec:iniGuess}), the specific definition of the piecewise constant approximation of the drift and diffusion functions (cf. section \ref{sec:iniGuess} as well), and the determination of a reasonable memory kernel length, which we regard as fixed during Bayesian estimation, based on the ACF of the data (cf. section \ref{choice}). The resulting technique offers, as we expect, various applications in the context of large data sets where memory effects are essential. As an example, we discuss an application from the field of turbulence (cf. section \ref{example}).

\section{Derivation of the estimation method}

\subsection{Model reformulation}\label{sec:reformulation}

To derive our method, we first propose an alternative formulation of the memory kernel via the term $\Delta X_{t,k} := X_t - X_{t-k\Delta t}$ which emphasizes the impact of trends: 
\begin{align}\label{modelGLE}
\dot{X}_t = D^{(1)}(X_t) + \sum_{k=1}^K \mathcal{K}_k\, \Delta X_{t,k} + \sqrt{D^{(2)}(X_t)}\,\eta_t.
\end{align}
The resulting model is equivalent to the original formulation of the GLE, since the term $\sum_{k=1}^K \mathcal{K}_k\, X_t$, by which the memory term has been extended here, can also be considered as part of the drift function, thus restoring the original form of the model. Through the focus on trends, the interpretation of the model equation is simplified. To explain this, let $X^{\ast}$ be a fixed point of the model belonging to a steady state $(x^{\ast}, x^{\ast}, ...)$. The value of the diffusion function $D^{(2)}(x^{\ast})$ defines the fluctuations around $X^{\ast}$. The memory term $\sum_{k=1}^K \mathcal{K}_k\, \Delta X_{t,k}$ equals zero at $X^{\ast}$, because we have $\Delta X_{t,k}=0$ in the steady state. Therefore, $x^{\ast}$ is a zero of the drift function: $D^{(1)}(x^{\ast})=0$. Conversely, a zero of the drift function defines a fixed point of the model. Thus, the GLE shares the equivalence between zeros of the drift function and fixed points of the model with the SLE. As a consequence, the formulation of the GLE based on trends allows for the use of an SLE estimate as an initial guess, which results in a very effective estimation of the GLE (cf. section \ref{sec:iniGuess}). Using the original memory kernel (cf. equation \ref{modelGLEoriginal}), this would not be possible.

Besides, we employ a time-discrete memory kernel, which suffices in the context of sampled measurements, and assume it to have a finite length $K$. The SLE corresponds to equation \ref{modelGLE} in the special case $K=0$ (we understand $\sum_{k=1}^0$ as a non-existent sum).

With regard to its complexity, the proposed version of the GLE is located between the SLE and the GLE with correlated noise. When we use the abbreviation ``GLE'' in the following, we refer to equation \ref{modelGLE}.

\subsection{Bayesian parameter estimation}\label{sec:Bayes}

In the sense of Bayesian estimation \cite{Toussaint,BayesianDataAnalysis} the posterior distribution $p(\vartheta|\xi)$ characterizes suitable model parameters $\vartheta=(\vartheta_1, ..., \vartheta_{N_{\text{par}}})$ based on the data $\xi$, which, in our case, is a measured time series $\xi=(x_0, x_1, ..., x_{N_{\text{data}}})$. Via Bayes' theorem, the posterior decomposes into likelihood $p(\xi | \vartheta)$ and prior $p(\vartheta)$ which are accessible quantities:
\begin{align}
p(\vartheta | \xi) \propto p(\xi | \vartheta) \, p(\vartheta). \label{Bayes}
\end{align}
The parameters $\vartheta$ of the GLE consist of the parameterization of the functions $D^{(1)}$ and $D^{(2)}$, which we will define later, and the values $\mathcal{K}_k$ of the time-discrete memory kernel. For now, we regard the length of the kernel $K$ as fixed and discuss its determination later (cf. section \ref{choice}).

Based on equation \ref{Bayes}, optimal model parameters can be identified by, e.g., the maximum (\textit{maximum a posteriori}, MAP) or the mean of the posterior. Alternatively, the whole posterior distribution $p(\vartheta|\xi)$ can be investigated via Markov chain Monte Carlo (MCMC) sampling to obtain complete information on possible parameter values. This is advantageous in the case of asymmetrical or bimodal distributions. In this work, we determine MAP estimates and calculate credible intervals based on the marginal distributions of the posterior. For a parameter $\vartheta_l$, the corresponding marginal distribution of the posterior is
\begin{align}
p(\vartheta_l|\xi) = \int\hdots\int p(\vartheta|\xi)\,d\vartheta_1 \hdots d\vartheta_{l-1}d\vartheta_{l+1}\hdots d\vartheta_{N_{\text{par}}}.
\end{align}
This integration is facilitated by the MCMC algorithm as a side product. To perform the MCMC computations we use an affine-invariant ensemble sampler implemented in the python package \textit{emcee} \cite{a:Foreman-Mackey2013}. Hereby, we use the MAP estimates to generate starting parameters. Regarding the optimization of the posterior which yields the MAP estimate, we have had good experience with a combination of the Powell \cite{Powell} and Nelder-Mead \cite{NelderMead} algorithms. The first is fast, the second is precise. Thus, applying the Nelder-Mead algorithm using the result of the Powell algorithm as an initial guess is a fast and precise choice.

\subsection{Constructing prior and likelihood}\label{sec:likelihood}

The prior $p(\vartheta)$ involves general knowledge about reasonable values of the parameters. Here we assign a probability of zero to all parameter values which violate the condition that $D^{(2)}$ must be a non-negative function.

As usual in the context of time series \cite{GreenlandIce,NARMAX,SLE_Likelihood_BIC,Bayes_SLE,Kleinhans}, the calculation of the likelihood $p(\xi|\vartheta)$ is performed in successive steps (neglecting the term $p(x_K, ..., x_0|\vartheta)$), each involving memory of depth $K$:
\begin{align}
p(\xi|\vartheta) = p(x_0, ..., x_{N_{\text{data}}} | \vartheta) = \prod_{i=K}^{{N_{\text{data}}}-1} p(x_{i+1} | x_i, ..., x_{i-K}, \vartheta).
\end{align}
The so called short-term propagator $p(x_{i+1} | x_i, ..., x_{i-K}, \vartheta)$ takes the form of a Gaussian distribution $\mathcal{N}(\mu,\sigma)$ where
\begin{subequations}
\begin{align}
\mu &= x_i + D^{(1)}(x_i)\,\Delta t + \sum_{k=1}^K \mathcal{K}_k\,(x_i - x_{i-k})\,\Delta t\\
\sigma &= D^{(2)}(x_i)\Delta t
\end{align}
\end{subequations}
\noindent This result can be derived through Euler-Maruyama approximation of the GLE \cite{GreenlandIce} (here, we assume that the sampling of the observed values is fine enough for this approximation of the estimated model to be valid). The latter reads (we use It\^{o} calculus) \cite{KloedenPlaten}
\begin{align}\label{EulerMaruyama}
X_{i+1} = X_i + D^{(1)}(X_i)\,\Delta t + \sum_{k=1}^K \mathcal{K}_k\,\Delta X_{i,k}\,\Delta t + \sqrt{D^{(2)}(X_i)}\,\sqrt{\Delta t}\, N_i
\end{align}
\noindent with $\Delta X_{i,k} := X_i - X_{i-k}$, $\Delta t$ being the time step, and $N_i\sim\mathcal{N}(0,1)$ being a sequence of independent Gaussian random numbers.

Due to technical reasons, one regards the logarithm of the posterior during both optimization and MCMC analysis. With the definitions $\mathcal{K}_0:=0$ and $\mathcal {K}_{-1}=\Delta t^{-1}$, we arrive at

\begin{align}\label{nll}
\log p(\vartheta|\xi) &= \log p(\xi|\vartheta) + \log p(\vartheta)\nonumber\\
&= \sum_{i=K}^{N_{\text{data}}-1} \Biggl\{ -\frac12 \log\left(2\pi D^{(2)}(x_i)\Delta t\right)\nonumber\\
 &\quad- \frac{\left( D^{(1)}(x_i)\Delta t + \sum_{k=-1}^K \mathcal{K}_k(x_i - x_{i-k})\Delta t \right)^2}{2D^{(2)}(x_i)\Delta t} \Biggr\} + \log p(\vartheta).
\end{align}

\subsection{Efficient realization of the estimation}\label{sec:efficient}

Performing the Bayesian estimation straightforwardly is numerically expensive for large data sets. During both numerical optimization and MCMC analysis, a large number of function calls of $\log p(\vartheta|\xi)$ have to be realized, taking the whole time series $\xi = (x_0, ..., x_{N_{\text{data}}})$ into account each time. In the following, we considerably raise the efficiency of the estimation by significantly reducing the number of floating-point operations (FLOPS) involved in a function call of $\log p(\vartheta|\xi)$. The underlying concept has previously been described for the SLE by Kleinhans \cite{KLEINHANS_MLE}.

First, as an approximation to arbitrary functions, we employ a piecewise constant parameterization of the functions $D^{(1)}$ and $D^{(2)}$:
\begin{equation}
D^{(1)}(x) = D^{(1)}_j,~ D^{(2)}(x) = D^{(2)}_j ~\text{for}~ x\in B_j.
\end{equation}
\noindent Hereby, we devide the range of the measured values $D = \lbrack x_{\min}, x_{\max} \rbrack$ into $N_{\text{bins}}$ parts $B_j$: $D = \cup_{j=1}^{N_{\text{bins}}} B_j$ (see section \ref{sec:iniGuess} for the choice of $N_{\text{bins}}$). Second, we perform the corresponding binning of the data and, third, we rearrange the summands in equation \ref{nll} accordingly by replacing the sum $\sum_{i=K}^{N_{\text{data}}-1}$ by the double sum $\sum_{j=1}^{N_{\text{bins}}} \sum_{x_i\in B_j}$. Using the following abbreviations
\begin{equation}
\Delta^k := x_i - x_{i-k},
\end{equation}
\begin{align}
c_j := \sum_{x_i\in B_j} 1, && c_j^{k} := \sum_{x_i\in B_j} \Delta^k, && c_j^{kl} := \sum_{x_i\in B_j} \Delta^k \Delta^l,
\end{align}
\begin{subequations}
\begin{align}
\beta^0_j &:= \frac12\log\left(2\pi D^{(2)}_j\Delta t\right) + \frac{(D^{(1)}_j)^2\Delta t}{2D^{(2)}_j} \\
\beta^1_{jk} &:= \frac{D^{(1)}_j}{D^{(2)}_j} \Delta t \mathcal{K}_k \\
\beta^2_{jkl} &:= \frac{1}{2D^{(2)}_j}\Delta t \mathcal{K}_k\mathcal{K}_l
\end{align}
\end{subequations}
\noindent we obtain after expanding the square in equation \ref{nll} and performing several reformulations:
\begin{align}\label{nll:sorted}
\log p(\vartheta|\xi) = -\sum_{j=1}^{N_{\text{bins}}} \Biggl\{ \beta^0_j c_j + \sum_{k=-1}^K \beta^1_{jk} c_j^k
+ \sum_{k=-1}^K\sum_{l=-1}^K \beta^2_{jkl}c_j^{kl}  \Biggr\} + \log p(\vartheta).
\end{align}

Here, the model parameters are to be found in the terms $\beta^0_j$, $\beta^1_{jk}$, and $\beta^2_{jkl}$ (as mentioned, we regard the kernel length $K$ as fixed at first, see section \ref{choice} for the choice of this value). The coefficients $c_j$, $c^{k}_j$, and $c^{kl}_j$ represent all information of the entire time series that is relevant to Bayesian estimation of the GLE (via its Euler-Maruyama approximation) and include the majority of FLOPS occuring in the unsorted expression of $\log p(\vartheta|\xi)$ (cf. equation \ref{nll}). Consequently, if the coefficients $c_j$, $c^{k}_j$, and $c^{kl}_j$ are evaluated only once before estimation, function calls of the sorted expression of $\log p(\vartheta|\xi)$ (cf. equation \ref{nll:sorted}) are several orders of magnitude cheaper than function calls of the unsorted one. To illustrate the proportions, in our example (cf. section \ref{example}) the lengths of the different sums occuring in the equations \ref{nll} and \ref{nll:sorted} are: about $10^7$ ($\sum_{i=K}^{N_{\text{data}}-1}$), typically $10^6$ ($\sum_{x_i\in B_j}$), 10 ($\sum_{j=1}^{N_{\text{bins}}}$), and up to 10 ($\sum_{k=-1}^K$). The analysis of large data sets, especially the computationally expensive MCMC sampling, is facilitated by this efficient procedure. It is worth mentioning again that this advantage can be utilized for arbitrary parameterizations of the drift and diffusion functions, if these are approximated piecewise constantly in every calculation of the posterior.

If necessary, the prior introduced in section \ref{sec:likelihood} could be extended by a term ensuring smooth curves in the piecewise constant parameterization of the drift and diffusion functions or the kernel function, as it is done in a different situation in reference \cite{Hummer_2005}.

\subsection{Initial guess}\label{sec:iniGuess}

Facilitated by our reformulation of the memory kernel based on trends, an estimate of the SLE provides an effective initial guess for the estimation of the GLE, further increasing the efficiency of the procedure (cf. section \ref{sec:reformulation}). We estimate the SLE employing the above posterior with $K=0$. Hereby we determine an appropriate distribution of the bins $B_j$. The number of bins $N_{\text{bins}}$ should meet the trade-off between flexible curves and arising statistical errors. To distinguish between valid details and statistical fluctuations of the curves, one can compare important statistical properties (such as the probability density function of the data points or the autocorrelation function) of the original time series and an integration of the model. The widths of the bins can be chosen uniformly or non-uniformly. An uneven choice should follow the distribution of the data points such that statistical errors are reduced. This so-called bandwidth selection has been discussed systematically in the Markovian case \cite{VarBandwidth1,VarBandwidth2}.

\subsection{Choice of the kernel length $K$}\label{choice}

During Bayesian estimation of the GLE model, we regard the kernel length $K$ as fixed. To identify a suitable value of $K$ covering all significant memory effects, a Bayesian model comparison could be employed. Here, often, the Bayesian information criterion (BIC) is used (as an approximation of the Bayes factor), which also avoids overfitting \cite{BIC, ABC}. We use a simpler procedure instead. We determine the kernel length by the curve of the function $\mathcal{K}_k$ which approaches zero in a physically meaningful model. The overall curve of $\mathcal{K}_k$ can be investigated by estimating kernel functions of different lengths $K$.

\begin{figure}
  \includegraphics[width=1\hsize]{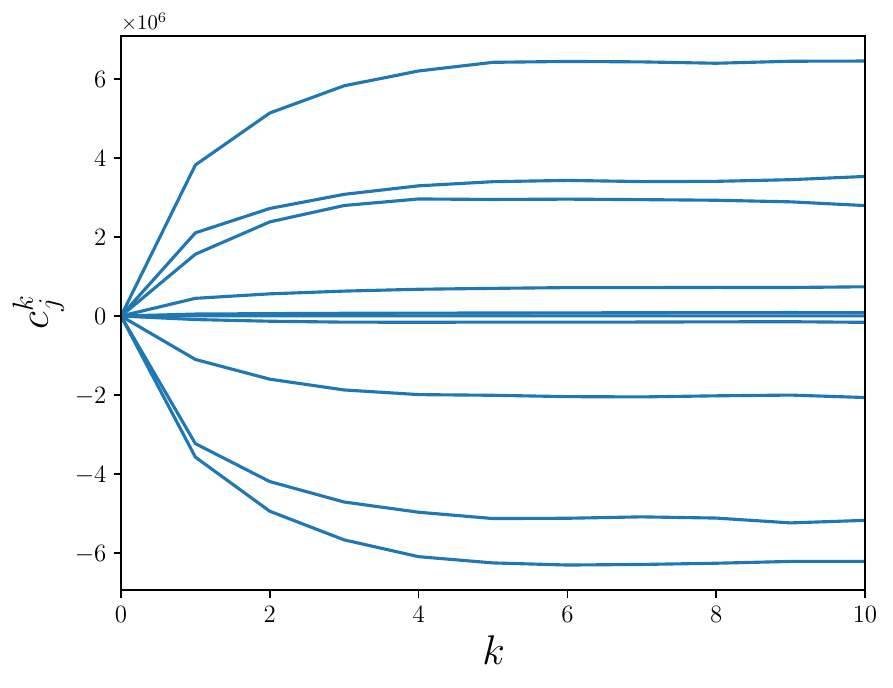}
  \caption{The coefficients $c_j^k$ approach plateaus for high values of $k$ (each line corresponds to a specific value of $j$). Memory effects of the trends $\Delta X_{t,k} = X_t - X_{t-k\Delta t}$ are significant before all these plateaus are reached. Underlying data set: free jet turbulence, $\tau=1400$.\label{fig:coeff}}
\end{figure}

Hereby, problems might arise if the chosen value of $K$ is too large, i.e., if the model is too different from the properties of the data. The coefficients $c_j^k$ typically show a monotonic behavior for increasing values of $k$ and reach constant values since they are bounded by the range of the data (cf. figure \ref{fig:coeff}). Information about the memory kernel is to be found in the change of the coefficients $c_j^k$. Once these are constant, the deterministic influence of the trends is hidden by the diffusive nature of the process and thus, the impact of the memory is no longer significant. If the chosen model includes memory terms for which all coefficients $c_j^k$ are constant, i.e., if it includes memory terms which are not relevant in the context of the data $\xi$ and about which the posterior provides no information, the estimation proves problematic.

Correspondingly, we propose to identify a limit $K^{\ast}$, such that Bayesian estimation of the GLE produces reliable results if $K<K^{\ast}$, by the point where the quantity
\begin{align}\label{kappa}
    \kappa_k := \sum_{j=1}^{N_{\text{bins}}} \left| c_j^k \right|
\end{align}
\noindent approaches its plateau. According to the above consideration as well as our experience with different examples, the kernel function always reaches values close to zero before $K^{\ast}$, i.e., significant memory effects are not to be found beyond $K^{\ast}$.

This issue is confirmed by the fact, that the point where $\kappa_k$ reaches its plateau corresponds to the point where the autocorrelation function (ACF) of the data $\xi$ approaches zero (cf. appendix \ref{appendix}). The ACF is usually employed as an indicator for dependencies in time. Here, it provides an upper limit for reasonable memory kernel lengths. An example is shown in figure \ref{fig:kappa}.

\section{Example: turbulent free jet data\label{example}}

Turbulent flows are ubiquitous in nature and technology. Due to their spatio-temporal complexity, modeling still remains a challenge. The general picture is that energy which is injected at large scales cascades towards small scales where it is dissipated \cite{Richardson,Frisch}. In a specific range in between, one assumes that the cascade process is affected  neither by the large scale boundary conditions nor by dissipation. This range is denoted as inertial range and bounded by two length scales: the \textit{correlation length} $L$ as upper bound and the \textit{Taylor length} $\Lambda$ as lower bound. Both length scales are based on the curve of the autocorrelation  function (ACF) $C(r) = (\langle u(r_0+r)u(r_0) \rangle - \langle u(r_0)\rangle^2)/\langle u(r_0)^2 \rangle$ of the longitudinal velocity $u(r)$, which therefore is an important statistical quantity of turbulent data. $u(r)$ is the result of a transformation of a measured time series $u(t)$ of the longitudinal velocity into a spatial velocity series by the Taylor hypothesis \cite{Turb:Reinke}.

Our goal is to apply our method to estimate a GLE model that reproduces the ACF of a free jet turbulence in air. Here, air streams out of a nozzle with diameter $d$ and the time series $u(t)$ is measured at a downstream position $z$ by, e.g., hot-wires. We use a data set from group (v) in reference \cite{Turb:Reinke} with a ratio of $z/d=40$ and a Taylor-Reynolds number of $Re_{\lambda}=865$. The time series contains $1.6\cdot 10^7$ data points, whereby the distance of two samples corresponds to a length of $1.31\cdot 10^{-4}\,\text{m}$. The unit of the measured velocities is $\frac{\text{m}}{\text{s}}$. For the sake of simplicity, we omit all units in the following. The Taylor length $\Lambda$ corresponds to roughly 50 samples, the correlation length $L$ to roughly 1450 samples.

As an example we analyze this time series by taking every 200th consecutive sample into account which is equivalent to a length scale that is located in the lower region of the inertial range. We use the parameter $\tau$ to denote the coarsening of the time series, such that we have $\tau=200$ in this case. Later we regard other length scales within the inertial range, i.e., other values of $\tau$.  We employ $N_{\text{bins}} = 10$ evenly distributed bins ($u_{\text{min}}=-7.27$, $u_{\text{max}}=10.31$). For the sake of simplicity, we choose $\Delta t = 1$ (a change of this value results in a scaling of the estimated parameters).

\begin{figure}
  \includegraphics[width=1\hsize]{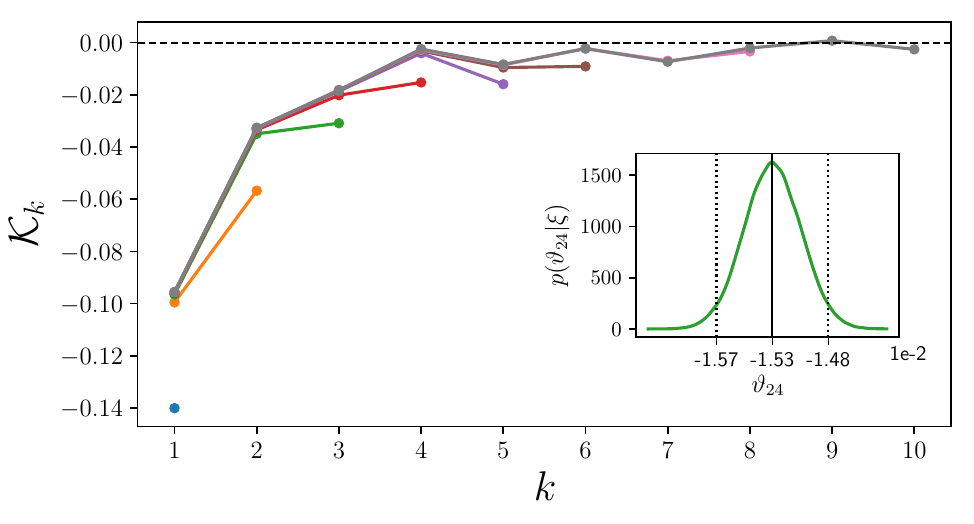}
  \caption{Values of estimated memory kernels $\mathcal{K}_k$ of the GLE for different kernel lengths $K$ in different colors. The credible intervals are smaller than the markers and not displayed for clarity. The inset shows an exemplary marginal posterior distribution belonging to the parameter $\vartheta_{24}$ (last kernel parameter for $K=4$) with estimates and a 95\% credible interval indicated by vertical lines. Underlying data set: free jet turbulence, $\tau=200$.  \label{fig:Kernels}}
\end{figure}

To determine a suitable memory kernel length $K$, we perform the estimation with different values and plot the resulting kernel functions (cf. figure~\ref{fig:Kernels}). All functions approach zero and show minor fluctuations for memory depths of $k=5$ and higher. They coincide with an overall curve, with the last value deviating from it. We interpret this as a compensation of the truncated deeper memory effects. Accordingly, we choose the estimated model belonging to the value $K=4$.

\begin{figure}
  \includegraphics[width=1\hsize]{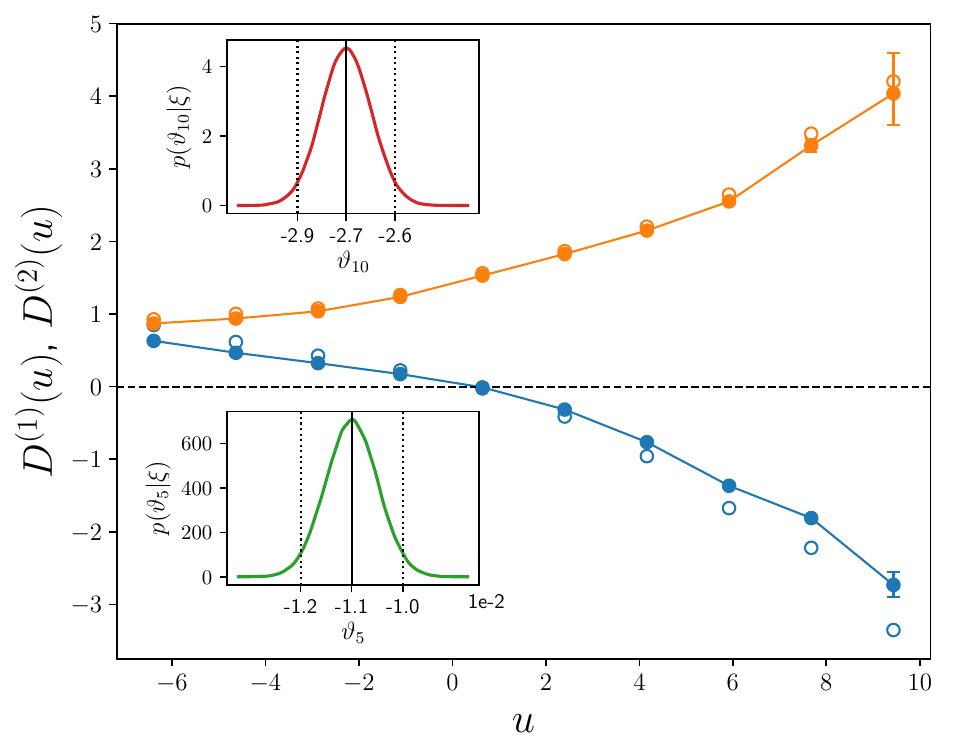}
  \caption{Values of the estimated functions $D^{(1)}$ (blue) and $D^{(2)}$ (orange) of the SLE (open circles; used as initial guess for the GLE) and the GLE (filled circles; $K=4$) at the centers of the bins with linear interpolations for the GLE result. According to their size, credible intervals are only visible at the rightmost values. The insets show exemplary marginal posterior distributions belonging to the parameters $\vartheta_5$ (zero of $D^{(1)}$) and $\vartheta_{10}$ (rightmost parameter of $D^{(1)}$) with estimates and 95\% credible intervals marked by vertical lines. Underlying data set: free jet turbulence, $\tau=200$. \label{fig:D12}}
\end{figure}

The estimated drift and diffusion functions are shown in figure~\ref{fig:D12} (filled circles). The initial guess (open circles) is adjusted only slightly. This indicates the efficacy of the proposed relationship between the SLE and the GLE based on the influence of trends (cf. equation \ref{modelGLE}). The essential advantage of the GLE model is not the adjustment of the drift and diffusion functions, but the introduction of the memory kernel.

Marginal posterior distributions and 95\% credible intervals are calculated by an MCMC analysis. Examples are plotted as insets and show clear and sharp distributions. Only in the rightmost bin, which contains less data points, the statistical errors are significant. Thereby, a credibility is always to be understood in relation to the given model framework. Regarding different models, a Bayesian model comparison can be carried out \cite{BIC}.

\begin{figure}
  \includegraphics[width=1\hsize]{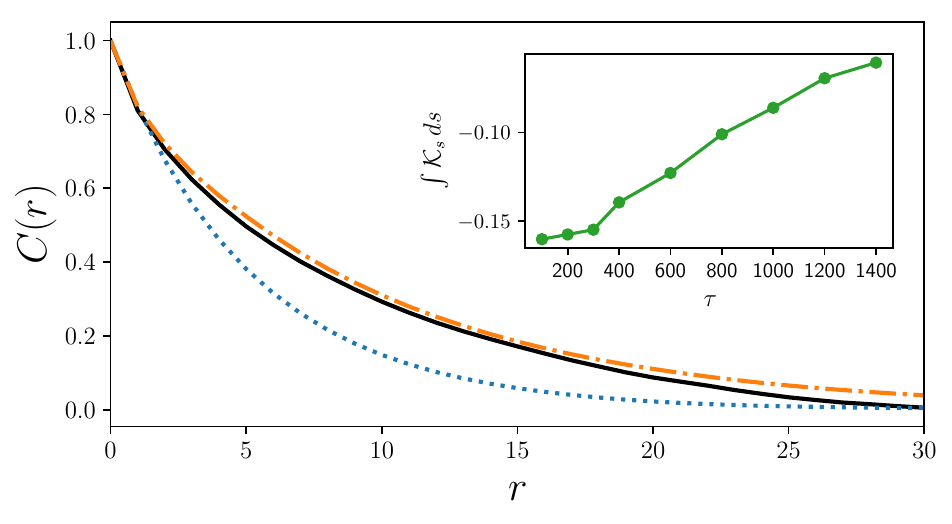}
  \caption{Autocorrelation function of original and simulated time series. Original data (black solid), SLE model (blue dotted), and GLE model (orange dash-dotted). The unit of the $r$-axis is the spacing between two consecutive samples. Underlying data set: free jet turbulence, $\tau=200$. \textit{Inset:} Integrals of estimated kernel functions for the free jet turbulence regarding different values of $\tau$ (coarsening of the sampling). The influence of the memory kernel approaches zero with $\tau$. \label{fig:ACF}}
\end{figure}

For the integration of an estimated model (cf. equation \ref{EulerMaruyama}), we interpolate the function values of $D^{(1)}$ and $D^{(2)}$ linearly between the centers of the bins (also higher order interpolations would be possible). Outside the interval $\lbrack u_{\text{min}}, u_{\text{max}}\rbrack$, we extrapolate $D^{(1)}$ and $D^{(2)}$ by constant values. The ACFs of the integrated time series of the estimated SLE and GLE are plotted in figure~\ref{fig:ACF}. The GLE leads to a good conformity with the original curve clearly outperforming the SLE. Hence, the consideration of memory effects significantly improves the modelling result.

When analyzing the data set on larger length scales, i.e., for higher values of $\tau$, the influence of the memory effects should decrease as the sampling of the original time series becomes coarser. To verify this relation, we plot the integrals of the estimated kernel functions (in the range $1\le k \le 5$) in the inset of figure~\ref{fig:ACF}.

\begin{figure}
  \includegraphics[width=1\hsize]{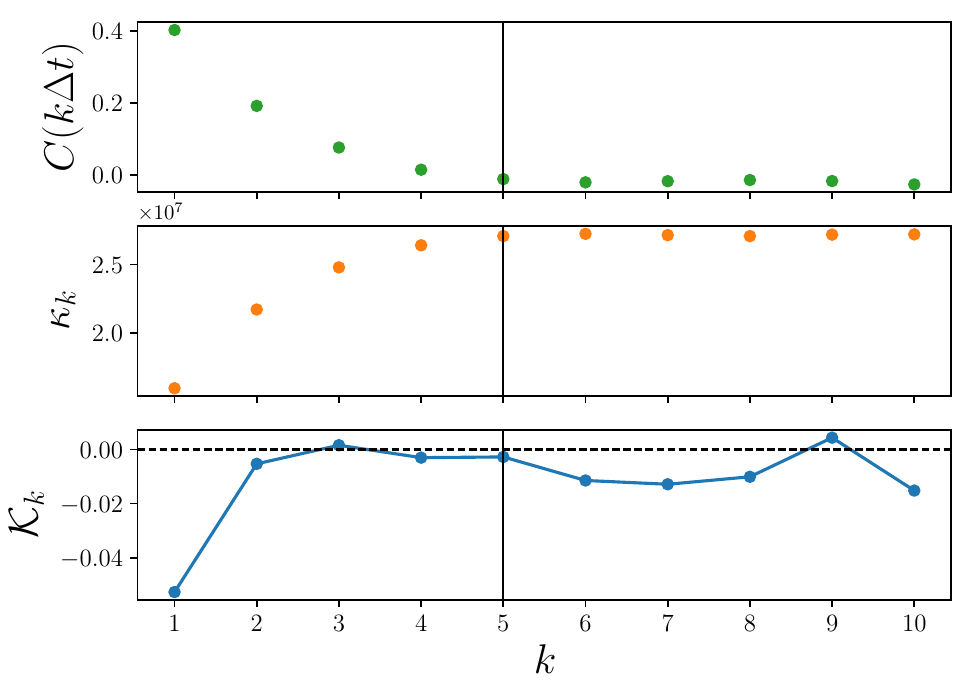}
  \caption{Estimated kernel function $\mathcal{K}_k$ (bottom). The corresponding plots of both the autocorrelation function $C$ and the quantity $\kappa_k$ (cf. equation \ref{kappa}) indicate that the estimation result is reliable up to $k=5$. Underlying data set: free jet turbulence, $\tau=1400$. \label{fig:kappa}}
\end{figure}

As explained in section \ref{choice}, the estimation of the kernel is reliable up to a specific value of $k$ which is determined by means of the autocorrelation function of the data or the quantity $\kappa$ (cf. equation \ref{kappa}). In our example, this is relevant for high values of $\tau$. For instance, for $\tau=1400$ the limit is $k=5$ as illustrated in figure~\ref{fig:kappa}.

\section{Conclusion and Outlook}

In this paper we have discussed an efficient implementation of Bayesian estimation of a GLE model from data. The chosen Bayesian approach facilitates the determination of both optimal parameters and credible intervals straightforwardly. The efficient implementation via a piecewise constant approximation of the drift and diffusion functions allows the analysis of large data sets. To be able to derive the likelihood directly via Euler-Maruyama approximation, we have regarded white noise. The application to turbulent free jet data has illustrated its capability in the context of experimental data of a system with memory. We expect that the version of the GLE proposed in equation \ref{modelGLE} also covers many other applications from different fields of science. As an example, we mention two fields of recent and future-oriented research in which the need for memory effects in modeling has come up in the last years: (paleo-)climate \cite{GreenlandIce,ElNino_DelayDE,Memory_Climate} and power grid \cite{Pesch, Katrin2DLangevin} modeling. Moreover, in these as well as in other fields, the anticipation of critical transitions is an important task. Thereby, recent studies show that common early-warning indicators, which rely on modeling by a (non-stationary) linear SLE (i.e., without memory effects), are likely to fail in realistic examples such as climate data \cite{Boers,Scheffer}.

The model allows for arbitrary, nonlinear drift and diffusion functions. Consequently, it further provides an extension to the intensively used autoregressive (AR) models which correspond to the time-discretized linear GLE \cite{GLE_PRE,linearGLE_ARMA}. It is also worth mentioning again that our method yields an estimate of the SLE in the case $K=0$ and, thus, provides a direct Bayesian extension of the method presented in reference \cite{KLEINHANS_MLE} (for Bayesian coarse graining in the Markovian case of SLE models, see also references \cite{Hummer_2005,Bayes_MD_Markov,Bayes_MD_Markov_2}).

Furthermore, in principle, the method can be extended to higher dimensional processes. Also the use of a nonlinear memory term $\sum_{k=1}^K\mathcal{K}_kf(\Delta X_{t,k})$ would  be conceivable, if the function $f$ is parameterized in a piecewise constant manner as well. Aside from disjoint bins, it may also be possible to define overlapping bins, such that the values of $D^{(1)}$ and $D^{(2)}$ are estimated by means of a moving window in space. This could be advantageous for very fine-structured functions. In a more general sense, kernel functions may be employed to avoid the discretization effect of the binning approach \cite{LangevinMLE_kernelDensity,NadarayaWatson}.

Regarding turbulence modeling it would also be of interest to apply our method to the analysis of the turbulent cascade in scale. In references \cite{Friedrich1997prl, Renner2001jfm} it was found that the dynamics of velocity increments between different scales can be described as Markov processes if the jump length between scales is larger than the Taylor length $\Lambda$. With our method this approach might be extended such that the description in scale is also possible for smaller jumps.\\

An exemplary python implementation of the proposed method is openly available \cite{Zenodo}.

\section*{Acknowledgments}

The authors thank J. Peinke and N. Reinke for providing the data set of the free jet turbulence measurement.

\appendix
\section{Appendix: Relation between $\kappa_k$ and ACF}\label{appendix}

In section \ref{choice} we stated that the point where the quantity $\kappa_k$ (cf. equation \ref{kappa}) approaches its plateau equals the point where the autocorrelation function (ACF) $C$ reaches zero. This relation holds since the convergence behavior of both quantities is determined by the term $\langle x_{i-k}|x_i\in B_j\rangle$ ($\lim_{k\to\infty}\langle x_{i-k}|x_i\in B_j\rangle = \langle x_i\rangle$):

\begin{align}
\kappa_k &= \sum_{j=1}^{N_{\text{bins}}} |c_j^k| = \sum_{j=1}^{N_{\text{bins}}} \bigl| N_j\, \langle x_i - x_{i-k} | x_i\in B_j \rangle \bigr| \nonumber \\
&= \sum_{j=1}^{N_{\text{bins}}} N_j\, \bigl| \bar{x}^j - \langle x_{i-k}|x_i\in B_j\rangle \bigr|,
\end{align}

\begin{align}\label{eq:app:C}
\langle x_i^2 \rangle\,C(k\Delta t) &= \Bigl\langle \bigl(x_i - \langle x_i\rangle\bigr)\bigl(x_{i-k} - \langle x_{i-k}\rangle\bigr) \Bigr\rangle \nonumber\\
&= \sum_{j=1}^{N_{\text{bins}}} \frac{N_j}{N_{\text{data}}} \Bigl\langle \bigl(x_i - \langle x_i\rangle\bigr)\bigl(x_{i-k} - \langle x_{i}\rangle\bigr) \Big| x_i\in B_j\Bigr\rangle \nonumber\\
&= \sum_{j=1}^{N_{\text{bins}}} \frac{N_j}{N_{\text{data}}} \Bigl\{ \langle x_i x_{i-k}| x_i\in B_j\rangle - \langle x_i\rangle\langle x_{i-k}|x_i\in B_j\rangle \Bigr\} \nonumber\\
&\qquad + \langle x_i\rangle \Bigl\{ \langle x_i\rangle - \sum_{j=1}^{N_{\text{bins}}} \frac{N_j}{N_{\text{data}}} \bar{x}^j \Bigr\} \nonumber\\
&\approx \sum_{j=1}^{N_{\text{bins}}} \frac{N_j}{N_{\text{data}}} \Bigl(\bar{x}^j - \langle x_i\rangle\Bigr) \langle x_{i-k}|x_i\in B_j\rangle.
\end{align}

\noindent Hereby, $N_j$ denotes the number of data points located in bin $B_j$ and $\bar{x}^j:=\langle x_i|x_i\in B_j\rangle$. The approximation $\langle x_i x_{i-k}| x_i\in B_j\rangle \approx \bar{x}^j \langle x_{i-k}|x_i\in B_j\rangle$ in equation \ref{eq:app:C} is valid, if the binning is fine enough.

\providecommand{\newblock}{}

\end{document}